\documentclass[final,5p,times]{elsarticle}

\usepackage{amssymb}

\usepackage{color}
\usepackage{listings}

\usepackage{color,hyperref}
\definecolor{darkblue}{rgb}{0.0,0.0,0.3}
\hypersetup{colorlinks,breaklinks,
            linkcolor=darkblue,urlcolor=darkblue,
            anchorcolor=darkblue,citecolor=darkblue}

\pdfpageattr {/Group << /S /Transparency /I true /CS /DeviceRGB>>} 

\journal{Astronomy \& Computing}

\begin{document}

\begin{frontmatter}


\title{SERPent: Automated reduction and RFI-mitigation software for e-MERLIN}



\author[label1]{Luke W.~Peck}
\ead{lwp@star.ucl.ac.uk}
\author[label1]{Danielle M.~Fenech}
\ead{dmf@star.ucl.ac.uk}

\address[label1]{Department of Physics \& Astronomy, University College London, Gower Street, London WC1E 6BT, UK}



\begin{abstract}

The Scripted E-merlin Rfi-mitigation PipelinE for iNTerferometry (SERPent) is an automated reduction and RFI-mitigation procedure utilising the SumThreshold methodology (Offringa et al. 2010b) \cite{Offringa_2010b}, originally developed for the LOFAR pipeline. SERPent is written in the Parseltongue language enabling interaction with the Astronomical Image Processing Software (AIPS) program. Moreover, SERPent is a simple `out of the box' Python script, which is easy to set up and is free of compilers. In addition to the flagging of RFI affected visibilities, the script also flags antenna zero-amplitude dropouts and Lovell telescope phase calibrator stationary scans inherent to the e-MERLIN system.

Both the flagging and computational performances of SERPent are presented here, for e-MERLIN commissioning datasets for both L-band (1.3 - 1.8 GHz) and C-band (4 - 8 GHz) observations. RFI typically amounts to $< 20 - 25\%$ for the more problematic L-band observations and $< 5 \%$ for the generally RFI quieter C-band. The level of RFI detection and flagging is more accurate and delicate than visual manual flagging, with the output immediately ready for AIPS calibration. SERPent is fully parallelised and has been tested on a range of computing systems. The current flagging rate is at 110 GB $day^{-1}$ on a `high-end' computer (16 CPUs, 100 GB memory) which amounts to $\sim$ 6.9 GB $CPU^{-1}$ $day^{-1}$, with an expected increase in performance when e-MERLIN has completed its commissioning.

The refining of automated reduction and calibration procedures is essential for the e-MERLIN legacy projects and future interferometers such as the SKA and the associated pathfinders (MeerKAT and ASKAP), where the vast data sizes ($>$ TB) make traditional astronomer interactions unfeasible.

\end{abstract}


\end{frontmatter}




\section{Introduction}
\label{sec:introduction}

Modern interferometers are becoming increasingly more sensitive and powerful, with resulting datasets becoming ever bigger. Therefore, the need for automation of certain procedures in reduction and calibration of interferometric data is vital. A major `bottleneck' in this reduction and calibration procedure is the manual removal of radio-frequency interference (RFI) and other bad unusable data by the user. Until recently, the manual flagging of typical datasets took a reasonable amount of time, with data sizes being on the order of Megabytes (MB). However, with improvements in receivers, electronics, correlators and optical fibre networks, observations now span a wide frequency range into bands, which are not protected for radio astronomy, thereby increasing the incidence of RFI. With future emphasis on multi-observation and full sky surveys (in the case of the SKA), data sizes will be on the order of Terabytes (TB), making manual flagging unfeasible. It is clear that automation of this process is necessary for the current generation of interferometers such as e-MERLIN, JVLA, ALMA, LOFAR and for future interferometers (MeerKAT, ASKAP, SKA).


One of the toughest challenges in RFI mitigation is accounting for its variable intensity, morphology and unpredictable nature. RFI can arise from many sources such as radio stations, microwaves, lightning, aeroplanes, mobile phones, CCTV etc. Some of these occur at specific frequencies (radio stations, mobile phones) and may only be problematic for certain arrays. Understanding individual array characteristics and the RFI environment it is situated in, needs to be considered to achieve optimal RFI reduction. Therefore, creating robust methods to mitigate RFI is essential. Mitigation can be applied at two stages in the interferometric data reduction process: pre-correlation and post-correlation and both can be complimentary to one another, as they will remove different kinds of RFI.


Pre-correlation is a very powerful option for RFI-mitigation because the observational data is still in its highest time resolution (sub-integration time) (Offringa et al. 2010b) \cite{Offringa_2010b}, although executing the processes on small sections of the entire observation at the station in real time is challenging. Numerous techniques have been developed for pre-correlation flagging, for example: thresholding methods using $\chi^{2}$ statistics (Weber et al. 1997) \cite{Weber_1997}, the cumulative sum method (Baan, Fridman \& Millenaar 2004) \cite{Baan_2004} and asynchronous pulse blanking (Niamsuwan, Johnson \& Ellingson 2005) \cite{Niamsuwan_2005}.


Post-correlation is the final stage to remove RFI before calibration procedures. Methods include the use of an independent RFI reference signal to subtract RFI (Briggs, Bell \& Kesteven 2000) \cite{Briggs_2000} and fringe-fitting for spatially and temporally constant RFI (Athreya 2009) \cite{Athreya_2009}. Thresholding methods remain an effective way to mitigate RFI as the amplitudes of the visibilities will be increased by the RFI. Offringa et al. (2010b) \cite{Offringa_2010b} analyse a number of threshold methods with simulated and real data from LOFAR and WSRT, and demonstrate that the SumThreshold method (explained in Section \ref{sec:threshold}) performs better than other rival methods. These include the Cumulative Sum method, VarThreshold and Singular Value Decomposition (SVD).

Since every interferometer around the world has a different baseline distribution, location, observed frequency band, RFI environment etc., the method of mitigation needs consideration and the implementation and parameters used may need to be optimised to suit any individual array. For example, WSRT is a large and sparse interferometer where the RFI is sometimes partially coherent. For this reason, post-correlation spatial processing algorithms are not always effective. Baan, Fridman \& Millenaar (2004) \cite{Baan_2004} conclude that real-time, pre-correlation time-frequency analysis conducted at each antenna would be more effective than any post-correlation method.

\begin{figure}[ht]
    \centering
        \includegraphics[scale=0.33]{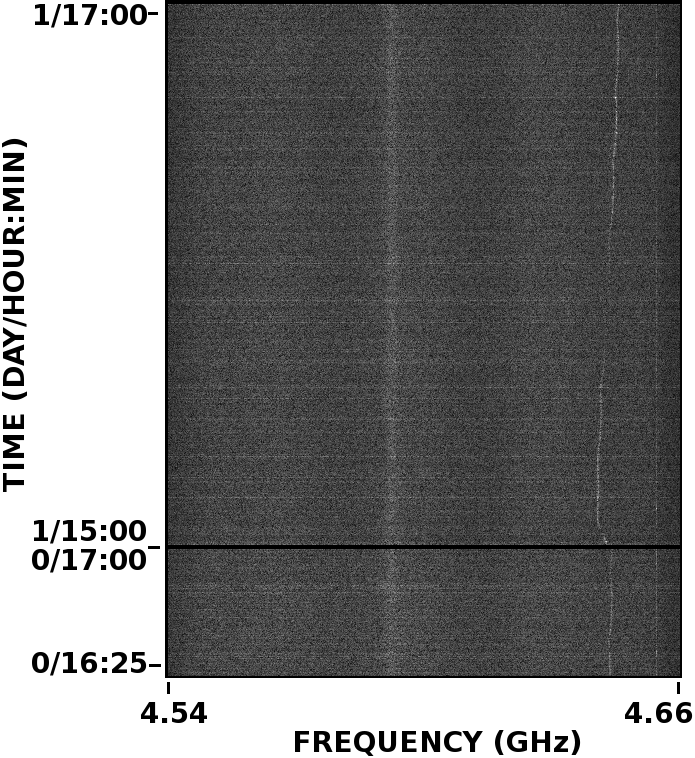} 
\caption{Time-frequency plot of the visibilities of the source 0555+398 from the COBRaS comissioning data W1 2011. A single IF and RR polarisation is shown with a frequency range from 4.54 to 4.66 GHz from the baseline Knockin-Pickmere ($5-7$). RFI is seen to vary both in time (vertical axes) and frequency (horizontal axes) at around 4.64 GHz.}
\label{fig:COBRaS_wiggly}
\end{figure}

In the case of e-MERLIN, there is no software or hardware available at Jodrell bank which would enable pre-correlation flagging to take place, other than any online flagging performed by the correlator. Therefore subsequent pipelines for e-MERLIN must include reduction and RFI mitigation processes which use post-correlation techniques such as those described by Offringa et al. 2010b \cite{Offringa_2010b}. Given the RFI environment for e-MERLIN and the incidence of RFI varying simultaneously over time and frequency (example: Figure \ref{fig:COBRaS_wiggly}), thresholding post-correlation methods are necessary in order to robustly mitigate this type of RFI.

Automated flaggers are compared on accuracy, computational performance, robustness and any technical requirements they impose (Offringa et al. 2010a) \cite{Offringa_2010a}. These criteria and the needs of the interferometer, will define which method is the most practical for that particular array. The aim of the Scripted E-MERLIN RFI-mitigation PypelinE for iNTerferometry (SERPent) is to provide an automated script which can be easily executed and combined with existing or future pipelines which fully reduces and flags radio interferometric data. This has been designed specifically for e-MERLIN and one of its Legacy projects: the Cygnus OB2 Radio Survey (COBRaS)\footnote{COBRaS: http://www.ucl.ac.uk/star/research/stars\_galaxies/cobras}, but is currently being tested on other instruments.



\begin{figure}[ht]
    \centering
        \includegraphics[scale=0.5]{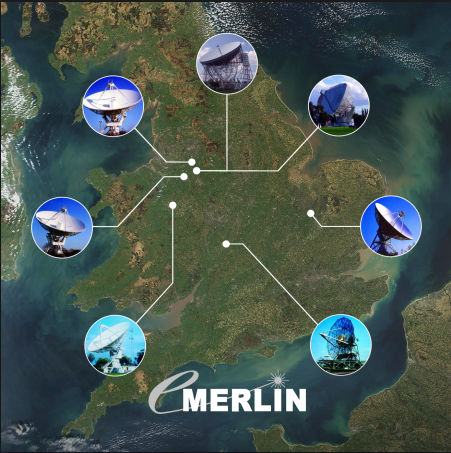} 
\caption{Positions of the seven radio telescopes of e-MERLIN across the United Kingdom. Clockwise from top; Lovell, Mark II, Cambridge, Defford, Knockin, Darnhall and Pickmere. The longest baseline is 217 km, giving resolutions of 150, 40 and 12 mas at 1.3 - 1.8, 4 - 8 and 22- 24 GHz respectively.}
\label{fig:emerlin}
\end{figure}

\section{e-MERLIN}
\label{sec:e-merlin}

e-MERLIN\footnote{e-MERLIN: http://www.merlin.ac.uk/e-merlin/} is a UK National Facility operated by The University of Manchester on behalf of the Science and Technology Facilities Council (STFC). It is an upgrade to the MERLIN (Multi-Element Radio Linked Interferometer Network) array, consisting of seven radio telescopes. Figure \ref{fig:emerlin} shows the distribution of telescopes spanning across the UK.

The upgrade consists of a new optical fibre network which connects each telescope to the Jodrell Bank Observatory, where the new WIDAR correlator developed by DRAO resides. New bandwidth receivers increase the useable bandwidth by two orders of magnitude, resulting in a continuum sensitivity increase of a factor of 10 or more compared to the old MERLIN array. There are three observing bands for e-MERLIN. L-band operates at 1.3 - 1.8 GHz, C-band at 4 - 8 GHz and K-band at 22 - 24 GHz, with the available maximum bandwidths of 512 MHz for L-band and 2048 MHz for C and K-bands per polarisation (circular).

In this paper, we will refer to observations with frequencies between 1.3 - 1.8 GHz as L-band as the bandwidth encompasses all of these frequencies. Observations with frequencies between 4 - 8 GHz will be referred to as C-band. All bands are comprised of smaller sub-bands or intermediate frequencies (IFs; in the AIPS nomenclature), which segregate the total bandwidth into groups of channels.


The shortest baseline of e-MERLIN is the Lovell - Mark II baseline of 400m, but the large difference in uv-spacing sampled between this baseline and the next shortest baseline of $\sim$ 11km (Mark II - Pickmere), means the Lovell - Mark II baseline is not used. This is because there is inadequate data to fully recover any diffuse structures seen on this very short baseline and connect the spatial scales detected by the other baselines in the array during the imaging process (Rob Beswick, private communication). Therefore, the smallest useable baseline is Mark II - Pickmere (11km) and the largest baseline is Lovell - Cambridge (217km).

This provides e-MERLIN with resolutions of 150, 40 and 12 mas for 1.3 - 1.8, 4 - 8 and 22- 24 GHz observations respectively. Table \ref{tab:e-merlin} gives the expected technical capabilities of a fully commissioned e-MERLIN array.

\begin{table}[ht!]
    \centering
    \caption{Technical Capabilities of e-MERLIN} \vspace{1mm}
    \begin{tabular}{l c c c}
    \hline\hline
                           & 1.5 GHz   & 5 GHz       &  22 GHz     \\
                           & (L-band)  & (C-band)    & (K-band)    \\ [0.5ex]
    \hline
    Resolution (mas)       & 150       & 40          & 12          \\ [0.5ex]

    Field of View (arcmin) & 30        & 7           & 2.0         \\ [0.5ex]

    Bandwidth (GHz)        & 0.5       & 2           & 2           \\ [0.5ex]

    Freq. Range (GHz)      & 1.3 - 1.8 & 4 - 8       & 22 - 24     \\ [0.5ex]

    Sensitivity ($\mu$Jy/bm)& 5 - 6    & 1.8 - 2.3   & $\sim$ 15   \\
    in full imaging run     &          &             &             \\ [0.5ex]

    Surface brightness      & $\sim$ 190& $\sim$ 70  & $\sim$ 530  \\
    sensitivity (K)         &          &             &             \\ [0.5ex]

    Astrometric             & $\sim$ 2 & $\sim$ 1    & $\sim$ 2    \\
    performance (mas)       &          &             &             \\ [0.5ex]

    Amplitude calibration   & 2\%      & 1\%         & 10\%        \\ [1ex]
    \hline
    \end{tabular}
    \label{tab:e-merlin}
    \newline
    \flushleft
    \vspace{-2.5mm}
    \footnotesize{General capabilities of the full e-MERLIN array. The sensitivity and surface brightness numbers include e-MERLIN and the Lovell telescope. The field of view decreases with inclusion of the Lovell telescope by approximately 20/(freq/ 1.4GHz) arcmin. This table is taken from the e-MERLIN website: http://www.e-merlin.ac.uk/tech/}
\end{table}

The new correlator at Jodrell Bank Observatory is a smaller version of the WIDAR correlator at the JVLA. A range of correlator capabilities are available for both continuum and spectral-line observations, and we refer the reader to the relevant literature for details (Garrington et al. 2004 \cite{Garrington_2004}, http://www.e-merlin.ac.uk/tech/).




The legacy projects and open proposals will feature a variety of astronomical themes and areas, requiring a slightly tailored approach to observational modes used for each proposal. This obviously has an effect on the data volume output from any observation. For a standard continuum observation in C-band from a full spec e-MERLIN, with a typical observing run of $\sim$ 18 hours in length the total data volume can be anywhere between 500 GB - 1 TB per day. This is assuming a typical time resolution of 1 second and frequency resolution of 250 kHz.

For spectral-line observations, the data volume can be expected to be even greater, with the highest available time and frequency resolutions of 1/4 seconds and 50 kHz respectively. All data is initially stored at Jodrell Bank and is copied and processed elsewhere.




\section{Post-correlation RFI-mitigation}

\subsection{SumThreshold Method}
\label{sec:threshold}

The current most effective thresholding method is demonstrated by Offringa et al. 2010b \cite{Offringa_2010b} to be the SumThreshold and this is the adopted RFI detection algorithm for SERPent. An overview of the method is given here, for a more in depth analysis of the process please see the afore-mentioned literature.


RFI increases visibility amplitudes for the times and frequencies they are present. Threshold methods work on the basis that if the RFI amplitudes are above a certain threshold condition, they are detected and flagged. The threshold level is dictated by the statistics of the relevant visibility subset, which can be the entire observation (all time scans, frequency channels, baselines etc.) or a smaller portion, for example: separate baselines, IFs and polarisations. This has the advantage of increasing the reliability of the statistics, because RFI may be independent of baseline and the distribution between IFs may differ. This is particularly relevant for L-band (1.3 - 1.8 GHz) observations where the RFI is more problematic.

Our tests also show that splitting the data in this manner, benefits the computational performance of SERPent. The reason for this is uncertain, however, this could be a function of memory usage within Python.


The SumThreshold method applied in SERPent, works on data which is separated by baselines and polarisations and arranged in a 2D array, with the individual time scans and frequency channels comprising the array axes i.e. time-frequency space. The frequency axis is further split by IFs due to the way the data is segregated within the fits file. The idea is that peak RFI and broadband RFI will be easily detectable when the visibility amplitudes are arranged in time-frequency space.

The e-MERLIN correlator outputs three numbers associated with any single visibility: the real part, the complex part and the weight of the visibility. When appending visibilities in the time-frequency space, if the weight is greater than 0.0 i.e. data exists for that time and frequency, then the magnitude of the real and complex part of the visibility is taken to constitute the amplitude. If the weight is 0.0 or less, i.e. no data exists for this baseline, time scan etc., then the amplitude is set to `NaN'. This has no effect on the statistics or threshold value, but acts as a substitute for that elemental position within the array, which both AIPS and SERPent require to retain the correct information. The Python module NumPy is employed to create and manipulate the 2D arrays, as the module is implemented with performance-optimized Fortran code\footnote{It should be noted here, that how this module is compiled can have a significant effect on performance.}.


There are two concepts associated with the SumThreshold method: the threshold and the subset size. A subset is a small slice of the total elements (in this case visibitility amplitudes) in a certain direction of the array (time or frequency). The difference between the SumThreshold method (a type of combinatorial thresholding) and normal thresholding is that after each individual element in the array has been tested against a threshold of N elements, the flagged values are averaged to the threshold level of subsequent runs. Moreover, the first threshold $\chi(1)$ is determined by statistics from the initial sample of visibilities, and subsequent thresholds $\chi(N)$ (where $N > 1$) are relative to $\chi(1)$. Threshold levels are discussed in more detail in section \ref{sec:estimators}.


Empirically a small subset $N = [1, 2, 4, 8, 16, 32, 64]$ works well (Offringa et al. 2010b) \cite{Offringa_2010b}. A window of size $N$ cycles through the array in one direction (e.g. time) for every possible permutation of connected samples for the given array and subset size. After each subset cycle a float array of identical size records the positions of any elements which are flagged. A 0.0 denotes a normal visibility, 1.0 signifies RFI in the time direction, 2.0 for the frequency direction and higher values for any subsquent runs of the flagger. At the beginning of the next subset cycle, for any element within the flag array whose value is greater than 0.0, the corresponding amplitude in the visibility array is reduced to the threshold level $\chi(N)$. If a group of elements of any subset size $N$ is found to be greater than the threshold level $\chi(N)$, then all elements within that window are flagged. This method is implemented in both array directions (i.e. time and frequency).

In addition to the SumThreshold methodology, certain clauses have been added to prevent the algorithm from over-flagging the dataset. If any threshold level reaches the $mean$ $+$ $variance$ $estimate$ the flagging run for that direction stops. The flagging process can run multiple times at the cost of computational time, and by default an initial run of subset $N = 1$ only, is included to remove extremely high amplitude RFI. This is followed by two full runs, providing that two conditions are met: the maximum value within the array after each run is a certain factor of the median and flags exist from the previous run. On each subsequent cycle, all flagged visibilities from the previous run are set to the next threshold in the visibility array so they don't skew the statistics and any weaker RFI which may remain can be found. This is necessary because some RFI in the e-MERLIN commissioning data are found to be over $10,000$ times stronger than the astronomical signal with some weaker RFI present.

For the first full run, the subset size doubles each step up to 32, and for the second full run, to $256$. This can be manually changed to lower values by the user to save time if there isn't much RFI in the observations.

\subsection{Statistical Variance Estimators}
\label{sec:estimators}

The variance of a sample is an important estimator of statistical outliers. Some statistical methods are sensitive to extreme values whereas others are robust against them. A study into a range of methods and various estimators are described and tested by Fridman (2008) \cite{Fridman_2008}. The median absolute deviation (MAD) and median of pairwise averaged squares are the most effective estimators that remove outliers, although Fridman (2008) \cite{Fridman_2008} comments that both are not as efficient, (i.e. needs more samples to produce the same power) as other methods. Since the sample size in any given observation from e-MERLIN will be of adequate size, this is not such an issue. The breakdown point for MAD is also very high (0.5), i.e. almost half the data may be contaminated by outliers (Fridman 2008) \cite{Fridman_2008}. MAD is adopted for this algorithm as an initial statistical estimator of the visibility population because of these robust properties. Again, Fridman (2008) \cite{Fridman_2008} stresses that the type and intensity of RFI, type of observation and the method of implementation are important factors when deciding what estimate to use for any given interferometer.


The MAD is the estimate of the variance used in the {\small SERPent} algorithm and is defined by Equation \ref{eq:MAD}, where $median_{i}(x_{i})$ is the median of the original population.
This median is then subtracted from every element in the population, creating a new modified sample of the same size as the original. The median of this new population is then calculated and multipled by a constant $1.4286$ to make this estimation consistent with that of an expected Guassian distribution.


\begin{equation}
MAD = 1.4826\; median_{j}\{|x_{j} - median_{i}\left(x_{i}\right)|\}
\label{eq:MAD}
\end{equation}

The first threshold level $\chi(1)$ is determined by the median of the sample (median($x_{i}$)), the variance estimator (MAD) and an aggressiveness parameter $\beta$ as shown in Equation \ref{eq:first_threshold} (Niamsuwan, Johnson \& Ellingson 2005) \cite{Niamsuwan_2005}. Since the median is less sensitive to outliers, it is preferred to the traditional mean in this equation and the MAD to the traditional standard deviation for the variance for similar reasons. If the data is Guassian in nature then the MAD value will be similar to the standard deviation (and the median to the mean). A range of values for $\beta$ has been tested for multiple observations and frequencies and a stable value of around $\beta = 25$ was found. Increasing the value of $\beta$ reduces the aggressiveness of the threshold and decreasing the value increases the aggressiveness.

\begin{equation}
\chi\left(1\right) = median_{i}\left(x_{i}\right) + \beta MAD
\label{eq:first_threshold}
\end{equation}

The subsequent threshold levels are determined by Equation \ref{eq:threshold} where N is the subset value, and $\rho = 1.5$, this empirically works well for the SumThreshold method (Offringa et al. 2010b) \cite{Offringa_2010b} and defines how coarse the difference in threshold levels is.

\begin{equation}
\chi(N) = \frac{\chi(N)}{\rho^{log_{2}\: N}}
\label{eq:threshold}
\end{equation}

\begin{figure*}[ht!]
    \centering
        \includegraphics[scale=0.2]{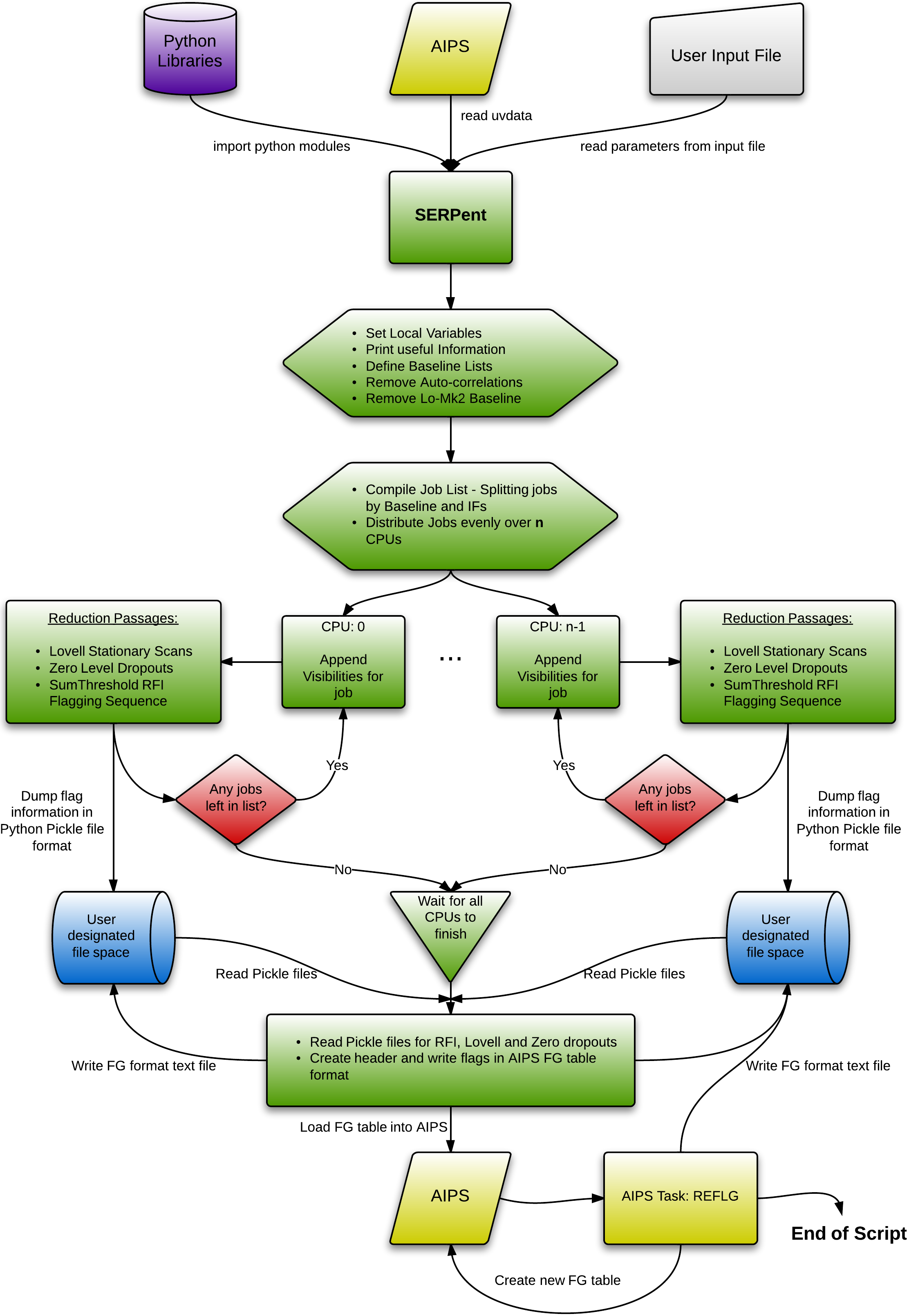} 
\caption{A logic flow chart of the SERPent process.}
\label{fig:flow_chart}
\end{figure*}

\subsection{SERPent Flagging Script}

It is anticipated that e-MERLIN data will be processed using a variety of software packages, most commonly AIPS (Greisen 2003 \cite{Greisen_2003}). Parseltongue\footnote{Parseltongue: http://www.radionet-eu.org/rnwiki/ParselTongue} is a Python based language which enables AIPS tasks to be imported as modules into the script. It is a popular choice for pipelines and is used extensively for European very long baseline interferometry (VLBI) network (EVN) calibration.\footnote{EVN Parseltongue pipeline:\\ http://www.jive.nl/wiki/doku.php?id=parseltongue:grimoire}.

SERPent is designed to be a simple, out of the box script. Hence, Parseltongue is an obvious choice because it has comprehensive access to AIPS tasks, and is independent of compilers as Python is the underlying language. For this reason SERPent is written in Parseltongue as opposed to a low-level language, despite the performance considerations which will be discussed in section \ref{sec:speed}.

\begin{table*}[htb!]
    \centering
    \caption{SERPent Performance Test Datasets} \vspace{1mm}
    \small
    \begin{tabular}{l c c c c c c c c c c}
    \hline\hline

    Dataset Name    &  Size    & Frequency  & Band &  Number of & Duration   & Baselines & Bandwidth   & IFs  &  Channels \\ 
                    &  (GB)    & Range (GHz) &      & Time-samples   & (Hours)     &           & (MHz)       &      &   per IF        \\ [0.5ex]
    \hline

    {\bf RFI Test Data:}  &         &       &     &      &   &       &               &     &  \\
          1436+6336       & 1.6   &  1.32 - 1.70     & L  & 5812  & 0.17       &  10       & 384    &  12    & 512   \\
          1407+284        & 432 MB   &  1.63 - 1.69 & L     & 73430 & 7      &  1        & 64     & 1      & 128  \\

    {\bf COBRaS W1 2011:} &        &        &    &     &        &         &       &      &  \\
          0555+398        & 2.3    & 4.41 - 4.92 &  C     & 99149  & 3       &  10       & 512     & 4     & 128    \\

    {\bf COBRaS 20$^{th}$ April 2012:}&           &      &           &                      &         &         &     &  \\
                     2033+4113        & 27.4     & 1.36 - 1.74 &  L     &  389839  & 20     &  21       &  384    & 12      & 128   \\

    {\bf COBRaS 18$^{th}$ July 2012:}& 97  & 5.49 - 6.00 &  C    &  1033940 & 26  & 21       & 512     & 4    & 512  \\
                 0555+398    & 10.5     & 5.49 - 6.00 &  C   &  112631  & 2  &  21      &  512    &  4   & 512  \\
                 1331+305    & 3.2     &  5.49 - 6.00 &  C     &  34124  & 0.5    &  21      &  512    &      4   & 512  \\
                 1407+284    & 3.4     &  5.49 - 6.00 &  C     &  35920 & 0.5     &   21     &  512    &      4   & 512   \\
                 2007+404    & 23.7     &  5.49 - 6.00 &  C     &  252745 & 12   &   21     & 512     &      4   & 512  \\
                 2032+411    & 56.1     &  5.49 - 6.00 &  C     & 598520  & 11   &  21      &  512    &      4   & 512  \\ [0.5ex]

    \hline
    \end{tabular}
    \label{tab:datasets}
    \newline
    \flushleft
    \vspace{-3mm}
    \footnotesize{Every dataset was observed with e-MERLIN with full circular polarisations (RR, LL, RL, LR). A list of associated sources has been provided here for each dataset.}
\end{table*}

Figure \ref{fig:flow_chart} shows a flow-diagram of SERPent to aid visualisation of the process. Green boxes represent SERPent processes, Yellow; AIPS, Purple; Python modules, Grey; User, Blue; Operating System and Red; Decision loops.



SERPent\footnote{SERPent software made publically available to download from the following location:\\ http://www.ucl.ac.uk/star/research/stars\_galaxies/cobras/technical/rfi} has now been tested on a number of systems and is stable.


\section{SERPent Results}

\label{sec:results}

Here we present various before and after plots demonstrating the reduction and flagging performance of the specific passages within SERPent. The auto-correlations are flagged by SERPent if present, along with the Lovell - Mark II baseline, because of the reasons stated in Section \ref{sec:e-merlin}. All datasets are listed in Table \ref{tab:datasets} and are commissioning e-MERLIN test datasets created purely to test pipelines and software such as SERPent or are commissioning Cygnus OB2 Radio Survey (COBRaS) datasets, with the view for first light on one of e-MERLIN's legacy projects.

\subsection{Lovell Stationary Scan Removal}
\label{sec:lovell}

A bright phase calibrator is observed for the technique of phase referencing, which is necessary for Very Long Baseline Interferometry (VLBI), in order to provide complex (amplitude and phase) solutions. This is achieved by alternating scans of the target and phase-cal source.

The Lovell telescope has a slow slew speed in comparison to the other telescopes within the array. This presents a unique problem to the e-MERLIN array. When phase-referencing it only participates in every alternative phase-cal scan, remaining stationary on the target for the other scans. This results in baselines containing the Lovell telescope to have two different amplitude levels for the phase calibrator.

In most cases the phase-cal will be brighter than the target source, thus when the Lovell is observing the phase-cal, the received flux will be greater than when the Lovell does not participate in the phase-cal scan and remains on the target source.

\begin{figure}[!htb]
    \centering
        \includegraphics[scale=0.35]{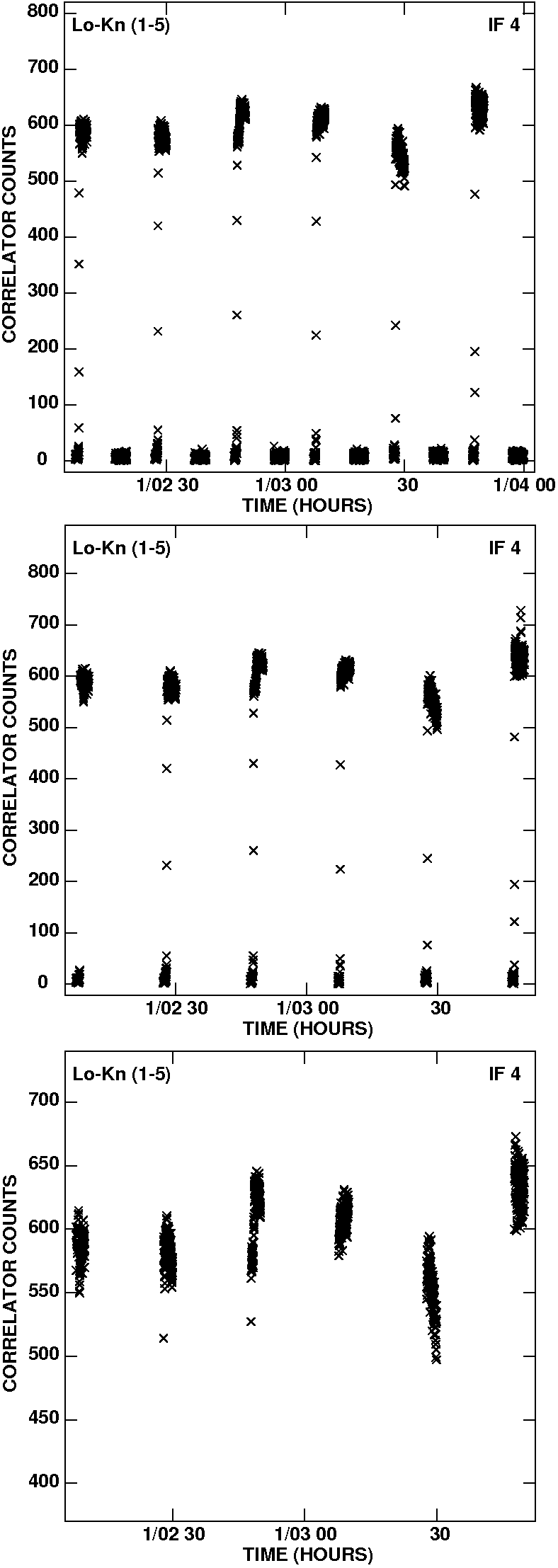} 
\caption{Amplitude-time plot with correlator counts on the y-axis and time on the x-axis, displaying a single IF and polarisation for the phase-cal source: 2007+404, baseline 1-5 (Lovell - Knockin) from the COBRaS July 2012 dataset. The top figure shows the visibilities before any flagging is done. The two distinct amplitude levels can be seen and the Lovell and zero-level dropouts are present. The middle figure shows the same visibilities after the Lovell stationary scan passage. The unconnected dropouts have all been flagged. The bottom figure shows the same visibilities after the Lovell and zero-level dropout passages. Both types of dropouts have been successfully flagged.}
\label{fig:lovell_zeros_combined}
\end{figure}

Figure \ref{fig:lovell_zeros_combined} shows the visibilities of the phase-cal for the Lovell-Knockin baseline, plotted in amplitude-time. The three windows display; top - before any flagging, middle - after flagging using the Lovell passage, and bottom - after flagging including the zero-level passage. There are two distinct amplitude levels, the highest is where the Lovell antenna contributes to the observation and the lowest is where the Lovell does not contribute.


SERPent detects whether the baseline contains the Lovell antenna and then executes the Lovell stationary scan passage appropriately. It defines each scan by checking whether the time duration between each scan is a factor larger than the integration time. If the average amplitude of all the visibilities within the scan is consistent with being a Stationary scan, it is flagged. This passage is essential for Lovell baselines. If the stationary scans (which make up 50\% of the total data) remain, the good phase-cal data would be treated as RFI in the flagging sequence and therefore flagged.


In Figure \ref{fig:lovell_zeros_combined} an additional effect can be seen which contributes to the zero-level amplitudes (see section \ref{sec:zero_level} for details). However, a careful inspection reveals that this additional zero-level contribution is part of the `on-target' Lovell scan, and not part of the Lovell stationary scan. The antenna has started to receive signal before the antenna has been properly alligned causing in-scan zero-level amplitudes to be observed. These are dealt with in another SERPent passage (see Section \ref{sec:zero_level} on zero-level dropouts).

SERPent's Lovell Stationary Scan passage removes the time intervals involved in the stationary scans for all channels within the tested IF from the NumPy visibility and flag arrays. A separate Lovell-only flag text file is created, as well as a combined master flag text file. This is done by dumping the flag information into a Python Pickle file, which is later read and combined with other files from other baselines and IFs. This combined flag text file is read into AIPS and attached to the input data as an AIPS flag extension table (FG), at the end of the script.



\begin{figure}[!htb]
    \centering
        \includegraphics[scale=0.35]{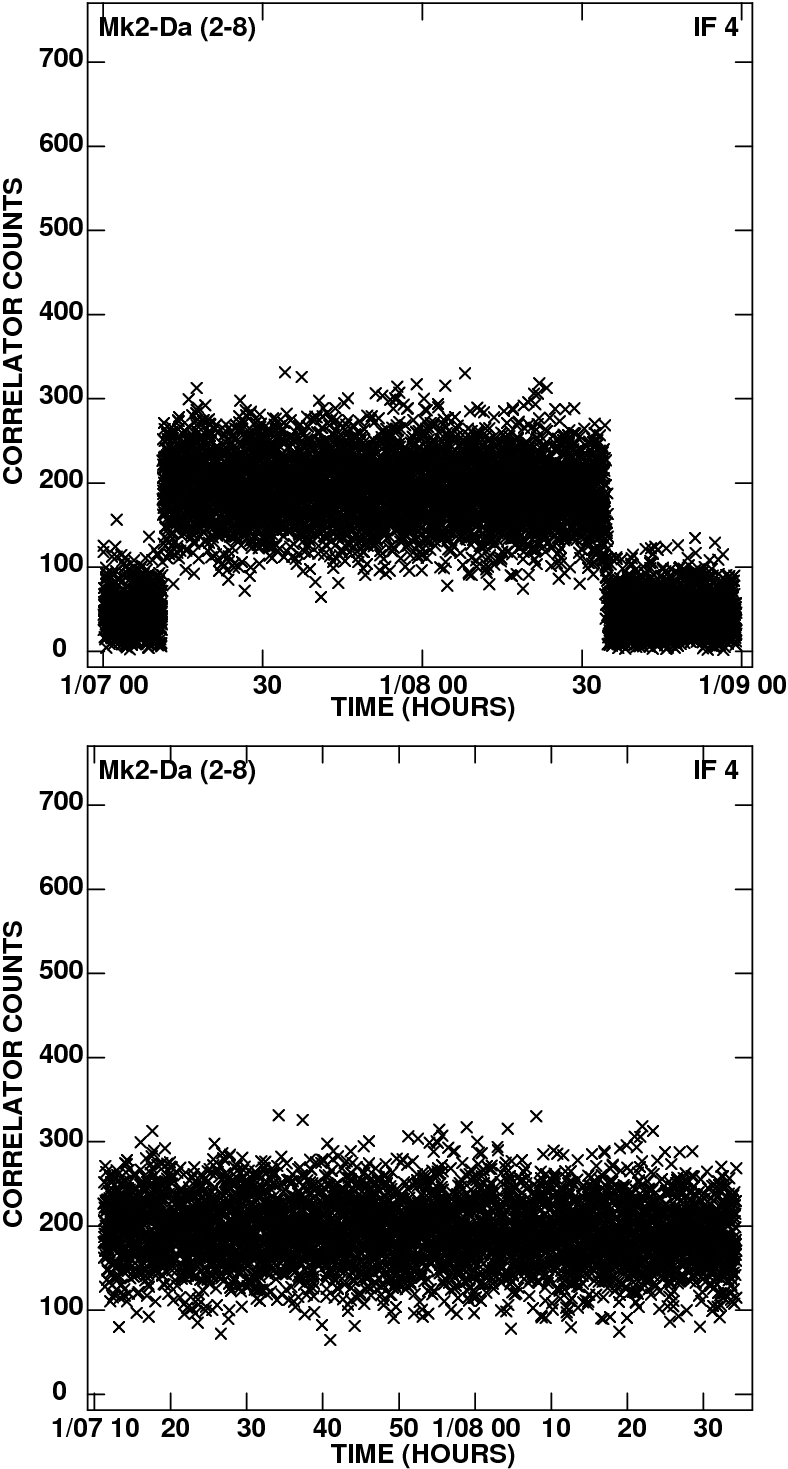} 
\caption{Amplitude-time plot with correlator counts on the y-axis and time in hours on the x-axis, of the source: 0555+398, baseline Mark II - Darnhall (2-8) from a single IF and polarisation from the COBRaS July 2012 dataset. The top figure shows the visibilities before the zero-level passage with two distinct amplitude levels at the beginning and end of the observation. The bottom figure shows the visibilities after the zero-level passage has been executed. The previous zero-level dropouts have been successfully removed.}
\label{fig:zeros_double}
\end{figure}

\subsection{Zero-Level In-scan Amplitude Dropouts}
\label{sec:zero_level}

Early COBRaS commissioning data revealed bad visibilities in the form of zero-level (visibility amplitudes of or around zero correlator counts) in-scan amplitudes (example: Figures \ref{fig:lovell_zeros_combined} and \ref{fig:zeros_double}), possibly a result of a system failure, telescope slew errors or the recording of data before the telescope was actually `on-source'. The zero-level amplitudes reside within scans containing good data and therefore need their own passage within SERPent to be flagged because these issues can arise on any baseline. This zero-level passage considers any visibility within all scans and it therefore does not matter where these zero-level amplitudes occur. It is expected that this effect will most likely occur either at the beginning or end of the scan.


%

\begin{figure*}[!htb]
    \centering
        \includegraphics[scale=0.35]{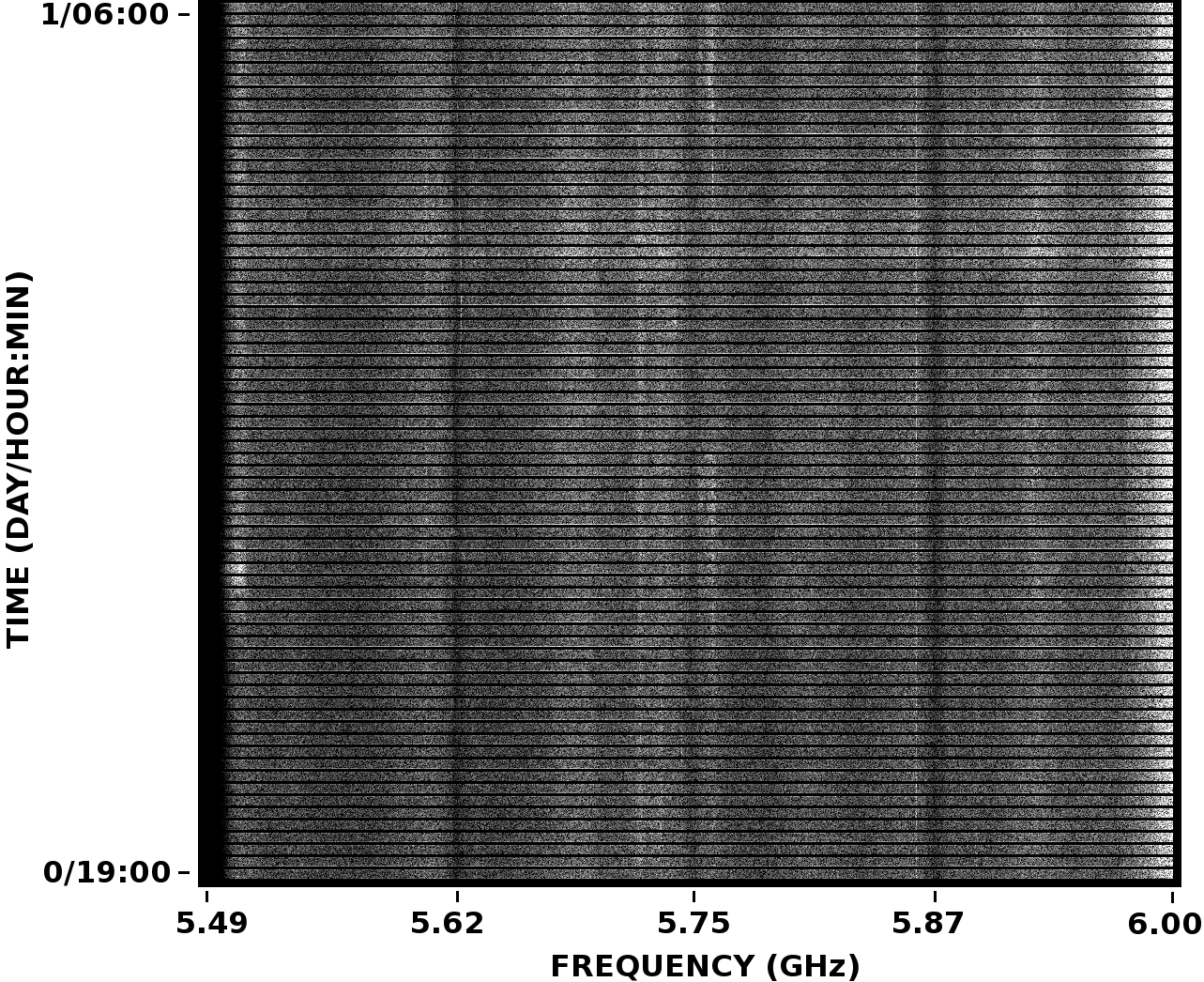} 
\caption{Time-frequency plot of the visibilities of the target field: 2032+411 from the COBRaS 18$^{th}$ July 2012 dataset. 4 IFs are plotted together with the bandwidth of 512 MHz from 5.49 to 6.00 GHz, from the Defford-Cambridge baseline. Weak narrowband and broadband RFI are present and noise in the edges of some IFs can also be seen.}
\label{fig:2032+411_130313_before}
\end{figure*}

In addition to the zero-level passage described above, it can be useful to trim the very edges of every scan, because SERPent can miss a few visibilities which are in transition between the zero-level dropout and on-source amplitude levels. The AIPS task QUACK can be implemented for this job for a very short section of the scan ($\sim 5s$), and is implemented in the COBRaS calibration pipeline after a full run of SERPent.

Figure \ref{fig:lovell_zeros_combined} shows the visibilities in an amplitude-time plot, after the Lovell (middle) and zero-level (bottom) passages have been performed. It can be seen that both have removed low-level off-source amplitudes which would have affected calibration and RFI-mitigation.

Another example of the zero-level in-scan amplitude dropouts can be seen in Figure \ref{fig:zeros_double} which shows the COBRaS July 2012 C-band dataset (source: 0555+398, baseline: Mark II - Darnhall (2-8)). This dataset contains zero-level dropouts at the beginning and end of the scan and also contains a few minutes of the previous source scan. This reinforces the idea that the zero-level dropouts result from telescope slews or from the correlation. Figure \ref{fig:zeros_double} demonstrates that SERPent's zero-level passage can deal with dropouts at the beginning or end of the scan after the successful flagging of these low amplitudes.



\subsection{RFI-mitigation Sequence}
\label{sec:RFI}

As discussed in Section \ref{sec:introduction}, RFI originates from a variety of sources. Some of the origins of RFI for e-MERLIN are known e.g. CCTV interference in L-band (1376 MHz), but others can be unpredictable, and neither are mitigated at the antenna or at the correlator level before data processing.



SERPent has been tested on both L-band (1.3 - 1.8 GHz) and C-band (4 - 8 GHz) observations (see Table \ref{tab:datasets} for datasets) which contain different amounts and types of RFI. L-band is typically noiser with both broadband and narrowband RFI common in observations, whereas C-band is generally RFI quiet with some narrow RFI common (though broadband RFI has been seen).

The edges of the IFs often contain noise as a result of the reduced response of the bandpass. SERPent can detect and flag this because it behaves in the same way RFI does. We now present a series of before and after figures which depict SERPent's flagging ability on a range of e-MERLIN datasets.

\begin{figure*}[!htb]
    \centering
        \includegraphics[scale=0.35]{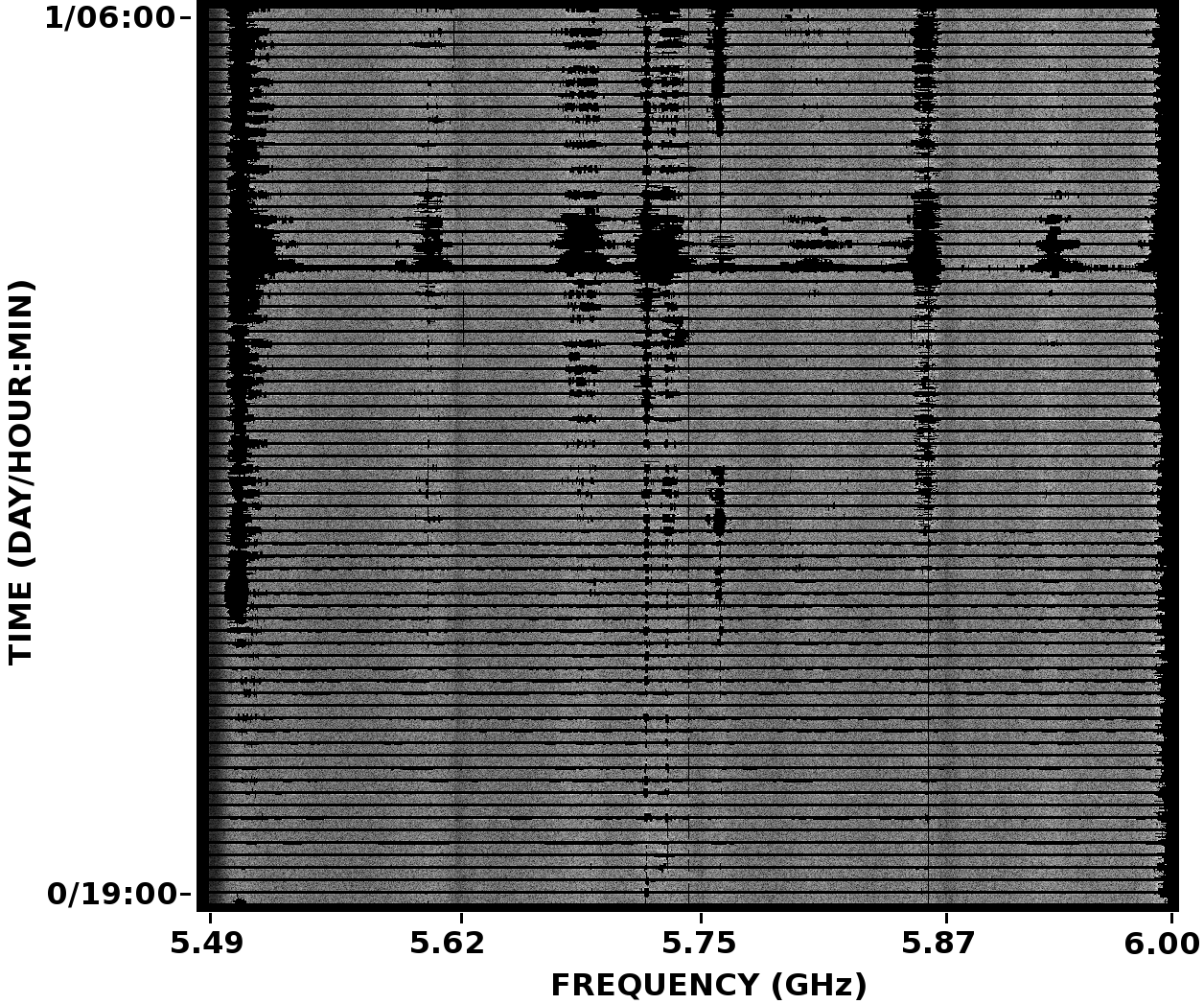} 
\caption{Time-frequency plot of the visibilities of the target field: 2032+411 from the COBRaS 18$^{th}$ July 2012 dataset. 4 IFs are plotted together with the bandwidth of 512 MHz from 5.49 to 6.00 GHz, from the Defford-Cambridge baseline. All of the visible narrowband and broadband RFI and the noise in the edges of IFs 1 and 4 has been flagged by SERPent.}
\label{fig:2032+411_130313_after}
\end{figure*}

Figure \ref{fig:2032+411_130313_before} displays the COBRaS C-band (centred on 5.49 GHz) 18$^{th}$ July 2012 data, with the visibilities sorted in time along the y-axis and channels in frequency along the x-axis, with all four IFs side by side. There is some weak narrowband and broadband RFI in the central channels and some noise present at the edges of IFs 1 and 4. Figure \ref{fig:2032+411_130313_after} shows the same data after SERPent flagging. All of the narrowband and broadband RFI and IF edge noise has been detected and successfully flagged. This level of RFI detection and flagging is more accurate and delicate than visual, manual flagging can achieve.



\begin{figure*}[!htb]
    \centering
        \includegraphics[scale=0.29]{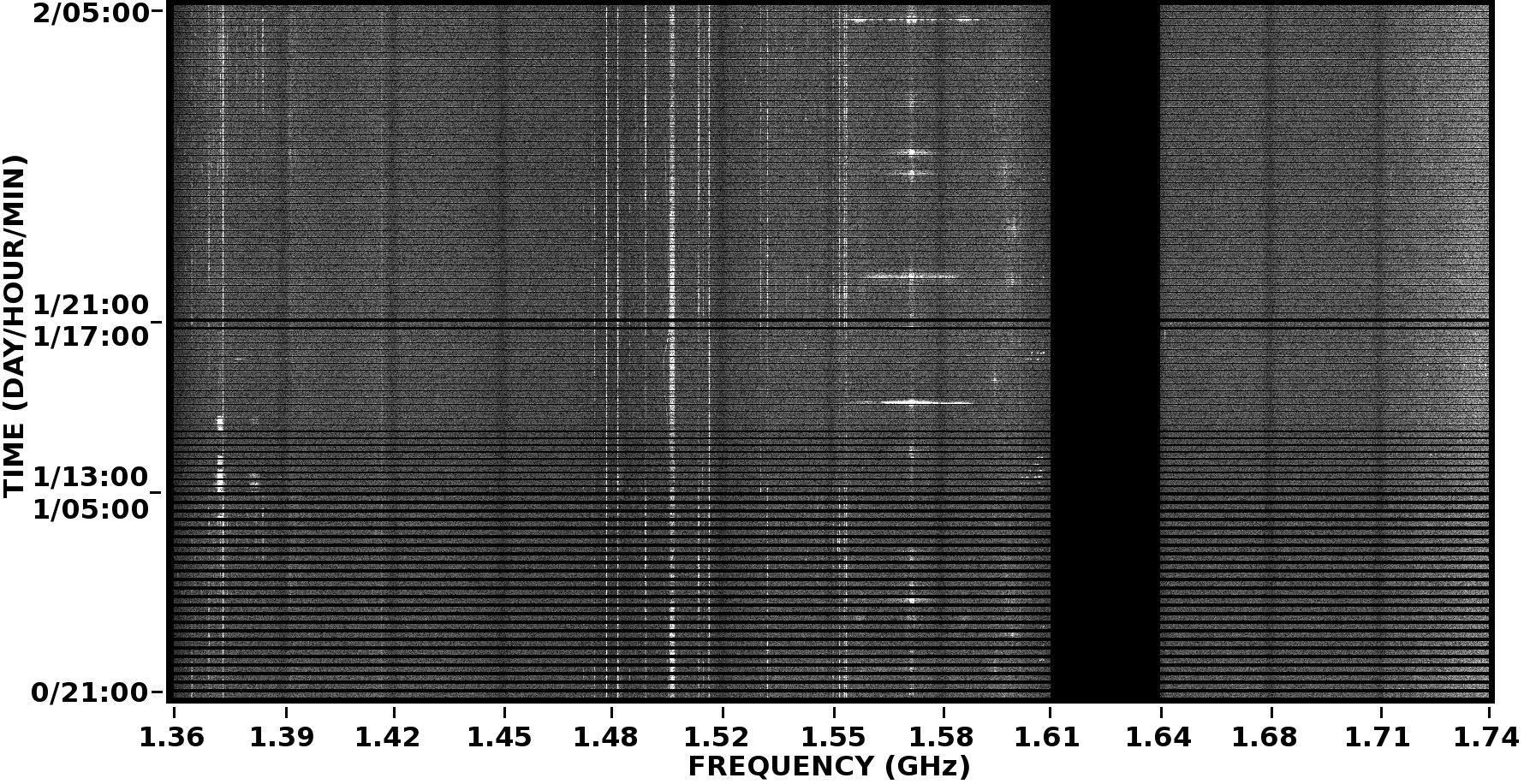} 
\caption{Time-frequency plot of the visibilities of the target field: 2033+411 from the COBRaS 20$^{th}$ April 2012 dataset. 12 IFs are plotted together with the bandwidth of 384 MHz from 1.36 to 1.74 GHz, from the Defford-Darnhall baseline. A variety of narrowband and broadband RFI can be seen, and many more weaker RFI are present but are below the current contrast levels, once the stronger visible RFI is removed, the weaker RFI is revealed. Note: IF 9 has been flagged by the online correlator before post-correlation reduction and processing.}
\label{fig:2033+411_lband_before}
\end{figure*}

The COBRaS 20$^{th}$ April 2012 L-band (centred on 1.56 GHz, with 12 IFs) datasets provide a greater test of SERPent's flagging capabilities because of the increased incidence of RFI. Once again the presence of narrowband and broadband RFI can be seen in Figure \ref{fig:2033+411_lband_before}. There is in fact more RFI present at lower levels, but this can not be seen in the spectral window before flagging. Note, IF 9 (1.61 - 1.64 GHz) has been automatically flagged by the correlator, before any processing of the data has been done.

\begin{figure*}[!htb]
    \centering
        \includegraphics[scale=0.29]{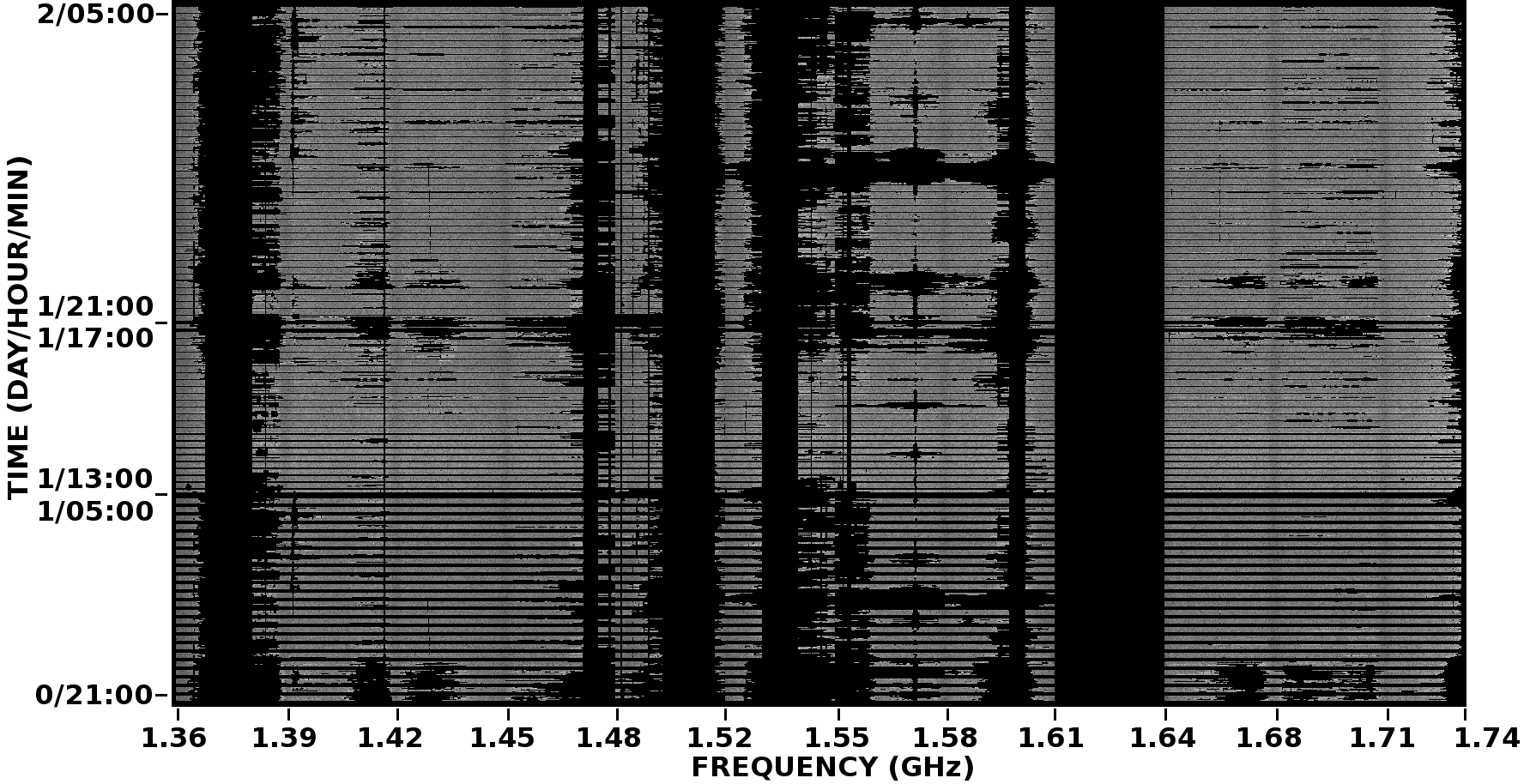} 
\caption{Time-frequency plot of the visibilities of the target field: 2033+411 from the COBRaS 20$^{th}$ April 2012 dataset. 12 IFs are plotted together with the bandwidth of 384 MHz from 1.36 to 1.74 GHz, from the Defford-Darnhall baseline. A lot of strong narrowband and broadband RFI has successfully been flagged, along with weaker RFI which was not visible in Figure \ref{fig:2033+411_lband_before}.}
\label{fig:2033+411_lband_after}
\end{figure*}

Figure \ref{fig:2033+411_lband_after} shows the L-band data following flagging by SERPent, again demonstrating the intricate nature of RFI detection by finding strong and weak RFI, as well as RFI which encompasses both large and small areas in the time-frequency space. Flagging to this level of accuracy on large datasets by hand would take an unfeasible amount of time.


\begin{figure}[!htb]
    \centering
        \includegraphics[scale=0.33]{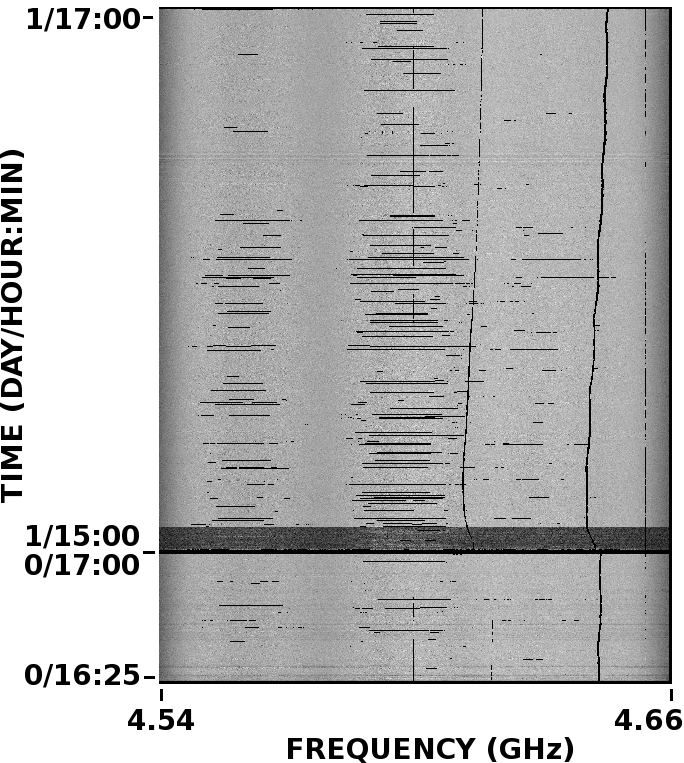} 
\caption{Time-frequency plot of the visibilities of the source: 0555+398 from the COBRaS W1 2011 dataset. A single IF and RR polarisation is shown with a frequency range from 4.54 to 4.66 GHz from the baseline Knockin-Pickmere. The before image can be seen in Figure \ref{fig:COBRaS_wiggly}. After a run of SERPent, the `wiggly' RFI has been flagged successfully.}
\label{fig:COBRaS_wiggly_after}
\end{figure}

There are examples of more exotic RFI in the commissioning datasets from e-MERLIN. The noisy COBRaS 2011 dataset at frequency 4.412GHz and source: 0555+398 shown in Figure \ref{fig:COBRaS_wiggly}, demonstrates some `wiggly' RFI which varied over time and frequency. As stated before, thresholding methods are the most robust way to detect these unusual types of RFI, and Figure \ref{fig:COBRaS_wiggly_after} displays how SERPent can deal with RFI of this nature.

\begin{figure}[!htb]
    \centering
        \includegraphics[scale=0.6]{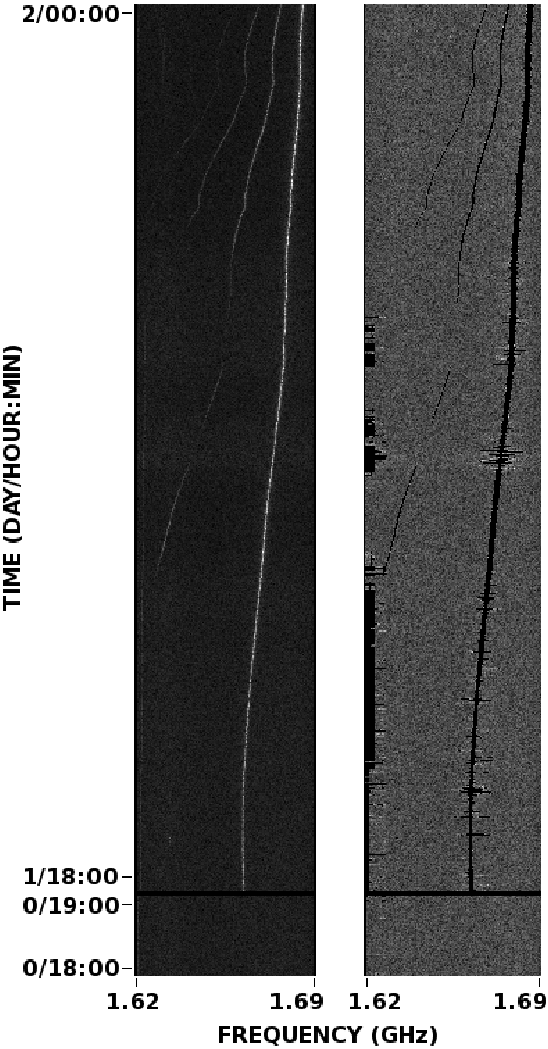} 
\caption{Time-frequency plot of the visibilities of the source: 1407+284. A single IF and RR polarisation is shown with a frequency range from 1.62 to 1.69 GHz from the baseline Lovell-Darnhall. Left; is the before image where the RFI varying in amplitude over time and frequency can be clearly seen, Right; is the clean, post-SERPent flagging image. Note that the contrast levels of the normal (unaffected) visibilities are different in each plot due to the influence of the RFI skewing the contrast levels.}
\label{fig:lightning}
\end{figure}

One further example of some peculiar multiple RFI found in e-MERLIN commissioning datasets can be seen from the source 1407+284, on the baseline 1-8 (Lovell - Darnhall) in Figure \ref{fig:lightning}. This RFI, of unknown origin, seems to drift in frequency over time and not necessarily in a constant direction. The before and after time-frequency plot in Figure \ref{fig:lightning} shows the complex shape of this RFI and how SERPent again has successfully flagged all of it.

These are only a small selection of examples from the commissioning e-MERLIN archives, but demonstrate the unpredictable nature of RFI and how thresholding detection methods can find RFI of any morphology. SERPent can easily convert this information into an AIPS readable FG table which is automatically appended to the input data in AIPS as part of the script.



\subsection{Computational Performance}
\label{sec:speed}

One of the important criterias for automated flaggers is computational performance. We have analysed the computational performance of SERPent on a number of computer systems, the details of which are given in Table \ref{tab:computers}. The difference in number of processors, Central Processing Units (CPUs) per processor and memory size covers a range of modest specifications available to institutions across the world (please refer to Table \ref{tab:computers} for details).


\begin{table}[ht!]
    \centering
    \caption{Computer Systems} \vspace{1mm}
    \begin{tabular}{l c c c}
    \hline\hline
    Computer Name    & Memory   & Processor       &  NCPUs     \\
                     & (GB)     &   (GHz)         &            \\ [0.5ex]
    \hline
    Leviathan (1 node)$^{1}$    & 100 & 3.20           & 16       \\
    Kria$^{2}$             & 40     & 2.93             & 24         \\
    Cornish1$^{2}$       & 16       & 3.20            & 8         \\
    Megan$^{3}$            & 48       & 2.40            & 16         \\
    \hline
    \end{tabular}
    \label{tab:computers}
    \newline
    \flushleft
    \vspace{-2.5mm}
    \footnotesize{Systems at: 1: University College London, UK. 2: University of Manchester, UK. 3: Netherlands Institute for Radio Astronomy (ASTRON).}
\end{table}



To increase computational performance, SERPent is parallelised by splitting the data into `jobs' which are evenly distributed across a number of CPUs. SERPent is parallelised in both baselines and IFs to maximize the even spread across CPUs and uses a user-designated number of CPUs specified in the input file. Our initial tests on modest data sizes reveal a significant increase in performance. These tests also show that the processing time scales linearly with the data volume.

\begin{figure}[!ht]
    \centering
        \includegraphics[scale=0.45]{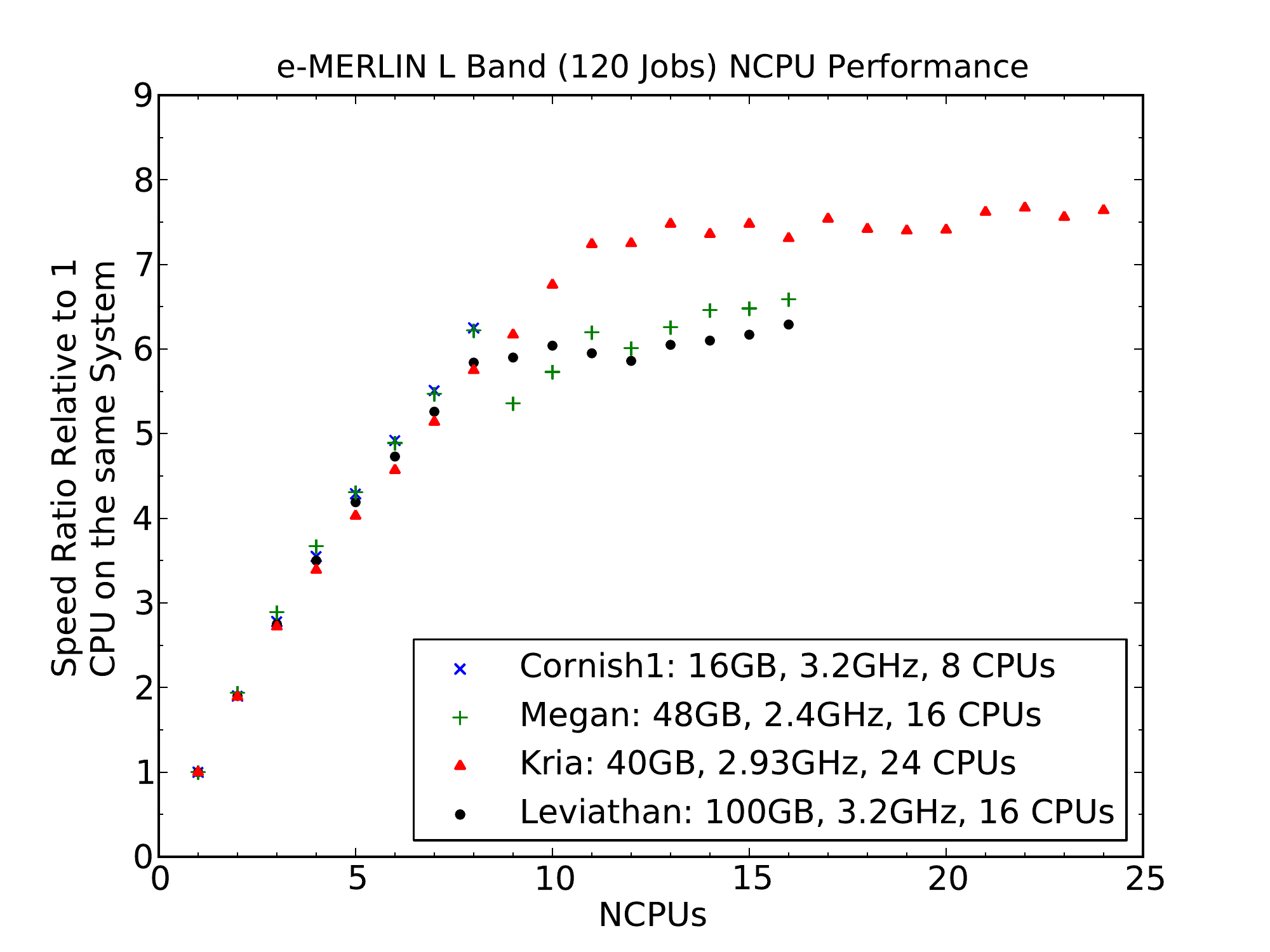} 
\caption{The speed relations of running SERPent on multiple CPUs on a range of computers relative to the performance of a single CPU on the same system. Even at high number of CPUs there are significant gains in performance which should increase further beyond 24 CPUs with datasets with a greater number of jobs available.}
\label{fig:SERPent_performance_plots_130313}
\end{figure}

Here we assess the effects of memory and the number of CPUs (NCPUs) used on the computational performance, by testing SERPent on a small dataset; RFI Test Data: 1436+6336, 1.63 GB (see Table \ref{tab:datasets} for details). Figure \ref{fig:SERPent_performance_plots_130313} portrays the relative performance ratio to a single CPU on the same system. All systems have linear relations with a peak CPU efficiency at around 8 CPUs. At this point adding more CPUs still increases performance but at a slower rate. We can infer that using 16 CPUs on this dataset has increased the performance by a factor $\sim$ 7 compared to using only 1 CPU on the same system. Runs on other datasets gave similar performance results.

The tail-off in Figure \ref{fig:SERPent_performance_plots_130313} may result from this particular dataset, where a few IF-baseline combos suffer from severe RFI (L-band observations). These jobs take more time to process, and as the number of jobs per CPU decreases, the portion of total time taken becomes biased towards the time taken by these `heavy' jobs. This is because all other CPUs have finished processing and are waiting idle whilst the CPU with the `heavy' job is still processing. Therefore, the performance relative to 1 CPU is affected by these jobs. However, the performance relative to 1 CPU is expected to increase further for a dataset with a higher number of total jobs (where the data has been distributed in a larger set of smaller segments), because the influence on the total computational time by any lengthy jobs is minimised. This, and the (random) distribution of jobs, is the reason a turnover after 8 CPUs is seen. This could also explain the fluctucations in performance on certain systems after 8 CPUs.

\begin{figure}[!ht]
    \centering
        \includegraphics[scale=0.45]{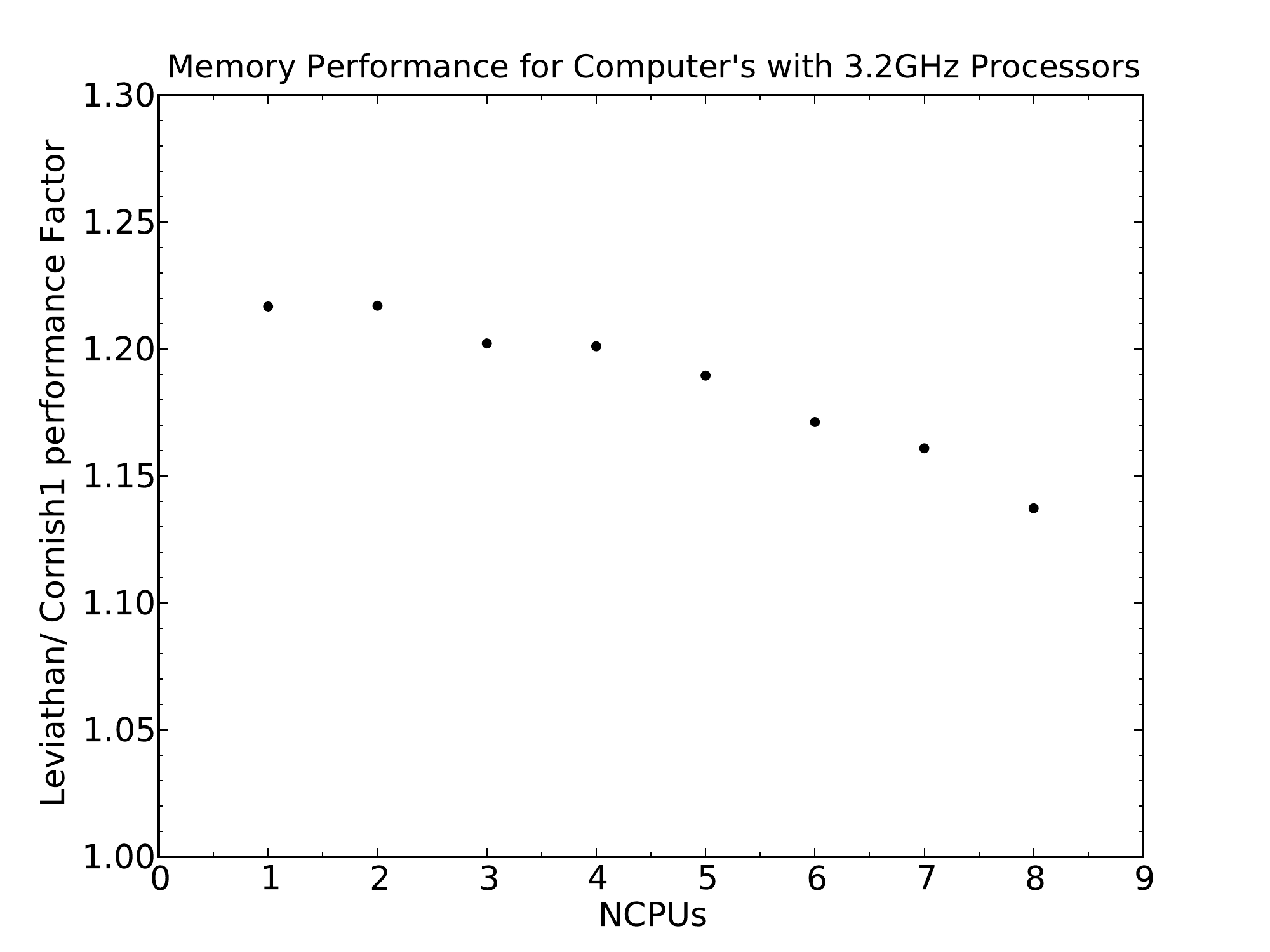} 
\caption{The performance relations of Leviathan/ Cornish1, both with 3.2GHz processors. As the number of CPUs increases the impact of Leviathan's larger memory per CPU on computational performance decreases.}
\label{fig:memory_performance}
\end{figure}

Increasing the amount of memory each CPU has also increased the computational performance, albeit by a smaller factor than the parallelisation. Comparing computers with the same processing speed (Leviathan and Cornish1 both have 3.2GHz processors), Leviathan has 6.25GB memory per CPU, and Cornish1 has 2GB memory per CPU. Figure \ref{fig:memory_performance} shows that the amount of memory per CPU decreases in significance as the NCPUs increases from a factor of 1.22 for 1 CPU to 1.14 when running on 8 CPUs. This is because the effect of parallelisation on performance is greater than the benefit of having extra memory per CPU. This shows that the limiting factor of running SERPent on interferometric datasets is the shear volume of data that needs processing over a number of CPUs and not the result of a lack of memory.

The raw (unaveraged) COBRaS July 2012 C-band data (97GB) takes 20 hours to process with SERPent, yielding a flagging rate of $\sim$ $110GB/day$. When the same dataset is averaged to 25GB, SERPent takes $\sim$ $6$ $hours$ with Leviathan (100GB Memory and 16 CPUs), which is consistent with linear scaling in data size and time. This approximately results in a processing rate of $6.9GB$ $CPU^{-1}$ $day^{-1}$. These extrapolations may vary in actual performance due to other factors such as the number of jobs SERPent creates, which is dependant on the number of baselines and IFs in any observation. The amount of RFI will also affect performance, as less RFI means SERPent can skip flagging runs due to kickout clauses in the flagging sequence etc. However, these remain reasonable estimates for predicted performances.


Full e-MERLIN Legacy observations will grow to data sizes of $\sim$ TB and contain up to 20 baselines (minus Lovell - Mark II baseline and autocorrelations) with (up to) 16 IFs for full 2 GHz bandwidth for C-band, and (up to) 12 IFs for full 512 MHz bandwidth for L-band. With these datasets SERPent will create 320 and 240 `jobs' for C and L-band respectively. It is clear from Figure \ref{fig:SERPent_performance_plots_130313}, that the plateau at around 7 results from the factorisation of the number of jobs (120 for this test dataset) and the NCPUs on each computer system. With the larger job list, it is expected that even larger increases in computational performances will be achieved.

To process a 1TB dataset in a day, the user will require $\sim$ $145$ CPUs processing at 6.9 GB $CPU^{-1}$ $day^{-1}$ (rate taken from processing COBRaS July 2012 C-band dataset). As discussed before, further increase in the number of CPUs won't result in an increase in performance because the job/CPU factor is the limitation in the parallelisation for the COBRaS July 2012 C-band dataset. However, an increase in the number of jobs because of an increase in the number of IFs for full e-MERLIN Legacy data will provide an increase on the factor 7 seen in Figure \ref{fig:SERPent_performance_plots_130313} from the parallelisation.

It would be simple to parallelise even further in polarisations, as currently every polarisation for each baseline and IF are contained within the same `job' but processed separately by the flagging sequence. This would potentially increase the number of jobs by a factor of 4 (for full polarisation studies). However, only computers with a high number of CPUs (NCPUs $>$ 100) would predominantly benefit from this, in addition to the increase in the number of jobs resulting from the increase in bandwidth.


In the scope of the COBRaS e-MERLIN legacy project (at University College London, P.I. Prof. Raman Prinja), the Leviathan computer system has 4 nodes (each with 100GB Memory and 16 CPUs), constituting a total of 64 CPUs. This enables $\sim$ $440$ $GB$ $day^{-1}$ of flagging to be processed (and possibly more with the increase in bandwidth and jobs), meaning a TB dataset would be processed in $\sim$ 2.25 days. This is a reasonable time scale for an automated script operating on a stable, full capacity instrument and a large survey such as COBRaS.


\section{Conclusions}
\label{sec:conclusions}

SERPent automates the reduction, flagging and preparation procedures of post-correlated radio interferometric datasets, specifically those from e-MERLIN. SERPent is in the process of being tested on EVN and Global VLBI datasets, showing good early results. This was done with Parseltongue, a common scripting language utilised prominently with the EVN, so that the user could starting flagging data which has been loaded within the AIPS environment with as little effort as possible. SERPent can be easily added to existing and future pipelines.

The entire SERPent package consists of only two text files: The main SERPent code to be executed, and a user input file, designed so the user does not have to interact with the main body of code, and so the input parameters are obvious and intuitive to set. This gives the freedom to the user to pursue their own flagging philosophy, i.e. whether they want to be aggressive or conservative with the flagging, but also includes a set of default inputs which will perform well on most datasets.

SERPent is designed to be run on `high-end desktop' computer systems. The examples in this paper used a system with 16 CPUs and 100GB of Memory (Leviathan) and was flagging at $\sim$ 110 GB $day^{-1}$. This throughput will increase with full e-MERLIN legacy data as the number of `jobs' will increase with full bandwidth, providing a higher throughput with a higher number of CPUs. It is unlikely that one will be able to process full e-MERLIN \emph{legacy} data on a modest desktop computer. Although obvious advantages in increased computer facilities and real world limitations on smaller systems are apparent, SERPent can be used by institutions without access to super computer clusters.

Section \ref{sec:results} has demonstrated that SERPent can reduce and flag current e-MERLIN commissioning data, which will have many more complications than a stable fully commissioned e-MERLIN including the Legacy datasets which will commence in the future. The benefit of using real data instead of simulated data is obvious, and SERPent is now part of the offical pipeline for e-MERLIN, used at Jodrell Bank and other international groups.


\section{Discussions}




When constructing an automated flagging script, the flagging philosophy has to be considered and decided. Whilst flagging all of the RFI and flagging none of the data is the idealistic scenario, even with implementing the SumThreshold Method with an extremely low false-positive detection percentage, either some RFI will remain or some good data will be flagged. This is the reality of working with real datasets from imperfect instruments and environments.

There are some philosophies which state `no data is better than bad data', promoting the more zealous and aggressive flagging, and others who would rather flag 80-90\% of RFI and have some of the weaker, lesser RFI remain. Obviously both philosophies can not be accommodated in total automation, therefore SERPent has the option for the user to decide some of the flagging parameters. These parameters include the aggressiveness, subset sizes and kickout thresholds. The AIPS REFLG task has also been seen to over-flag at times, although it is necessary to condense the number of rows in the AIPS FG table.


The computational performance of SERPent is probably the area which requires most improvement. It currently flags $\sim 110GB/day$ with 16 CPUs, which is reasonable for commissioning e-MERLIN datasets. However, for the fully upgraded e-MERLIN this will be slow. It is obvious that including more CPUs could solve this problem, as 16 CPUs is still very modest in modern computing terms, however this is merely shifting the problem onto hardware (and isn't very constructive). The flagging sequence makes two full passes through the SumThreshold method (the original AOflagger Offringa et al. 2010 \cite{Offringa_2010b} makes 5 passes) in order to maximise RFI detections, and skips these passes if the threshold level is low enough. This is currently the limiting factor in terms of performance. Reducing this to one full pass would speed SERPent up considerably at the expense of RFI-mitigation performance. Note that the amount of RFI also affects computational performance, because more RFI means more full runs completed within SERPent, and less RFI means more cycles are skipped due to the invoked kickout clauses implemented in the flagging sequence to stop over flagging and increase speed performance.

Comparing SERPent with flagging implementations on the JVLA and LOFAR, the data volume per processing time appears to be slower. In the case with LOFAR, the AOflagger has been written in a low-level language (C++) and includes specific compiler settings to achieve the optimal performance \cite{Offringa_2012}. In addition, the AOflagger is heavily parallelised over multiple cores and nodes on a super cluster, vectorised, and is part of the LOFAR pipeline which fully reduces and calibrates observations for users. This is different in the case of e-MERLIN, where the data will still be in a raw format when presented to the user, who will not have access to the same computing facilities as LOFAR. There is work currently being conducted on a general e-MERLIN pipeline, and SERPent is the flagging software implemented for the reduction passage. However, this is only a general pipeline and does not account for the many calibration techniques and methods needed for the many diverse projects e-MERLIN will observe for.

In the case of the JVLA, there is no implementation that is as sophiscated in mitigating RFI as the AOflagger or SERPent methods. The CASA software package is the main choice for the JVLA, and all developments are focused to this package. On the contrary, e-MERLIN currently favours AIPS because of the fringe fitting abilities within the program needed to calibrate e-MERLIN data.

According to received feedback, SERPent can be rather aggressive at times. Whilst differing flagging philosophies can account for these views, it should also be considered that e-MERLIN is still a commissioning interferometer and thus a changing and unstable system. Every dataset is unique but can also represent e-MERLIN in a new commissioning stage. We personally have experienced filter issues with some of the COBRaS April 2012 L-band datasets which have since been resolved, but caused amplitude level issues which then affected RFI-mitigation performance. This is the nature of commissioning instruments, particularly in the case of e-MERLIN, a heterogeneous array whose antennas have other responsibilities outside of e-MERLIN (Lovell and Cambridge partake in EVN observations). Compared to other, dedicated arrays such as the VLA/ JVLA and ALMA, both homogeneous (ALMA has 2 types of antennas) arrays which have been modelled extensively before commissioning. This provides a much smoother transition from the commissioning to fully-commissioned phases for the JVLA and ALMA. These factors should not be over looked with respect to e-MERLIN commissioning, because both hardware and software changes make maintaining external software such as SERPent difficult.


Furthermore the tweaking of SERPent flagging parameters may still yet yield the most optimised settings for both flagging and speed performances. The best time to conduct and hone these settings will be once e-MERLIN has settled and finished its commissioning phase.


As with all practical software, SERPent has limitations. The only limitation in detecting Lovell Stationary scans (Section \ref{sec:lovell}) would be when the phase calibrator is weak and of a similar amplitude level to the dropout. Since the code finds individual scans and makes comparisons with the previous scan this should still function correctly. The zero-level dropout passage (Section \ref{sec:zero_level}) also has obvious limitations. In the COBRaS July 2012 C-band dataset the Cambridge antenna stopped collecting data during $\sim 75\%$ of the observation of the target source and phase calibrator (phase referencing mode). Whilst this represents similar morphology to the zero-level dropout (which are mostly slew/ timing errors) when three quarters of the data is bad, it is hard to automatically flag these visibilities without comprising the rest of the data. It is reasonable to assume that a fully stable e-MERLIN system will provide data without such problems as a minimum requirement of data quality assurance.

Lastly, we discuss the limitations with the RFI-mitigation sequence (Section \ref{sec:RFI}) in SERPent. Setting the number of full SumThreshold flagging runs to two will increase flagging accuracy at the cost of speed as discussed earlier. In terms of flagging limitations, SERPent has a limit of $\sim 45\%$ of the sample population being RFI (or statistical outliers) for any one SumThreshold run. This results from using the Median Absolute Deviation (Section \ref{sec:estimators}) as a robust estimate of variance (Fridman 2008 \cite{Fridman_2008}). Hence, the amount of RFI SERPent can deal with depends on the strength of RFI and how many SumThreshold runs are implemented. For example, if 40\% of RFI exists for the first full SumThreshold run and another 40\% exists in the remaining sample at a lower amplitude level, SERPent can theoretically deal with a total of 60\% of RFI in the beginning sample. This is dependent on the RFI being represented at multiple amplitude levels, and with two full SumThreshold flagging runs. Even more RFI could be mitigated, if more full runs are implemented.


SERPent has yet to be tested on spectral-line observations or pulsar/ transient observations. The methodology of SERPent would suggest that any spectral-line would risk being identified as RFI, depending on the full width half maximum (FWHM) and/ or strength of the line profile. If the frequency and redshift of the spectral line is known, then it is possible to create a mask for frequencies where the spectral line resides. These frequencies would then be protected from the main RFI flagging sequence and perhaps have their own flagging passage to remove RFI which may populate the same frequencies as the spectral-line.


There is much scope for future work and improvements to SERPent and automated scripts for radio interferometry, with the SKA and its pathfinders (MeerKAT and ASKAP) on the horizon. As the datasets become larger, the necessity of automated scripts to perform the majority of the reduction and calibration becomes essential. SERPent currently interacts with AIPS by reading the visibilities into Python NumPy arrays and then creating the FG flag tables which are read back into AIPS. Since all of the flagging and processing is conducted within the NumPy environment, it is possible that SERPent could be adapted to work with CASA (a more recently developed software package for radio astronomy data reduction) datasets with minor adaptations.

For SERPent, any future work will be concentrated on stabilising the flagging parameters for a stable e-MERLIN system, and improving the computational performance.


\section{Acknowledgements}

e-MERLIN is a national facility operated by The University of Manchester on behalf of the Science and Technology Facilities Council (STFC). LP wishes to thank supervision from Prof. Raman Prinja and Dr. Jeremy Yates. LP and DF both wish to acknowledge funding from STFC. Parseltongue was developed in the context of the ALBUS project, which has benefited from research funding from the European Community's sixth Framework Programme under RadioNet R113CT 2003 5058187.

We acknowledge and thank; Ivan Marti-Vidal and Eskil Varenius (Chalmers University of Technology, Sweden) for their parallelisation code which the parallelisation in SERPent is adapted from. Rob Beswick (Manchester, UK) for providing commissioning e-MERLIN RFI test data and for testing the performance of SERPent on the computers at The University of Manchester. Megan Argo (Netherlands Institute for Radio Astronomy ASTRON) for testing the performance of SERPent on the computers at ASTRON and general feedback. Robert Laing (ESO Garching, Germany) for feedback with problems and issues which have helped the development of SERPent. We thank all the people who discussed and commented on SERPent in the RadioNet Advanced radio interferometry, commissioning skills and preparation for the SKA workshop (Manchester, UK, November 2012).

We would also like to thank the anonymous referee, who provided extensive detailed comments and corrections.





\bibliographystyle{elsarticle-harv}
\bibliography{SERPent_ref} 







\end{document}